\documentclass[prl,twocolumn,groupedaddress]{revtex4}
\usepackage{graphics,epsfig}

\begin{document}

\title{Abrupt structural transitions involving functionally optimal networks} 
\author{Timothy C. Jarrett$^+$, Douglas J. Ashton$^*$, Mark Fricker$^{**}$ and Neil F. Johnson$^+$}

\affiliation{$^+$Physics Department, Oxford University, Oxford, OX1 3PU, U.K.}

\affiliation{$^*$Physics Department, University of Nottingham, Nottingham, NG7 2RD, U.K.}

\affiliation{$^{**}$Department of Plant Sciences, Oxford University, Oxford, OX1 3RB, U.K.}

\date{\today}

\begin{abstract}
We show analytically that abrupt structural transitions can arise in functionally optimal networks, driven by small changes in the level of transport congestion. Our findings are based on an exactly solvable model system which mimics a variety of biological and social networks. 
Our results offer an explanation as to why such diverse sets of network structures arise in Nature (e.g. fungi) under essentially the same environmental conditions. As a by-product of this work, we introduce a novel renormalization scheme involving `cost motifs' which describes analytically the average shortest path across multiple-ring-and-hub networks.

\end{abstract}

\maketitle

There is much interest in the {\em structure} of the complex networks which are observed 
throughout the
natural, biological and social sciences \cite{newman,newman2,watts98,nets,nets2,charges,central,search,gradient,prl,dm}. The physics community,
in particular, hopes that a certain universality might exist among such networks. On the other hand, the biological community knows all too well that a wide diversity of structural forms can arise under very similar environmental conditions. In medicine, cancer tumors found growing in a given organ can have very different vascular networks. In plant biology, branching
networks of plant roots or aerial shoots from different species can co-exist in
very similar environments, yet look remarkably different in terms of their structure.  
Mycelial fungi \cite{mark} provide a particularly good example, as can be seen in Figs. 1(a) and 1(b) which show different species of fungi forming networks with varying degrees of lateral connections (anastomoses).
Not only do fungi represent a wide class of anastomosing, predominantly planar, transport
networks, but they have close parallels in other domains, including vascular networks,
road and rail transport systems, river networks and manufacturing supply
chains. But given that such biological systems could adapt their structures over time in order to optimize their functional properties, why do we observe such different structures as shown in Figs. 1(a)  and 1(b) under essentially the same environmental conditions?

Here we address this question by showing that quite different network structures can indeed share very
similar values of the functional characteristics relevant to growth. We also show analytically that small changes in the level of network congestion can induce abrupt changes in the optimal network structure. In addition to the theoretical interest of such phase-like structural transitions, our results suggest that a natural diversity of network structures should arise under essentially the same environmental conditions -- as is indeed observed for systems such as fungi (see Figs. 1(a) and (b)). As a by-product of this analysis, we provide a novel renormalization scheme for calculating analytically the average shortest path in multiple ring structures.

Our analytically-solvable model system is inspired by the transport properties of real fungi (see Fig. 1(c)). A primary functional property of an organism such as a fungus, is to distribute nutrients efficiently around its network structure in order to survive. Indeed, fungi need to transport food (carbon (C), nitrogen
    (N) and phosphorous (P)) efficiently from a localized
source encountered on their perimeter across the structure to other parts of
the organism. In the absence of any transport congestion effects, the average shortest path would be through the center -- however
the fungus faces the possibility of `food congestion' in the central region since the mycelial tubes carrying the food do not have infinite capacity.  Hence the organism must
somehow `decide' how many pathways to build to the center in order to ensure nutrients get passed across the structure in a reasonably short time. In other words, the fungus -- either in real-time or as a result of evolutionary forces -- chooses a particular connectivity to the central hub. But why should different fungi (Figs. 1(a) and (b)) choose such different solutions under essentially the same environmental conditions? Which one is `right'? Here we will show that, surprisingly, both fungi can be right at the same time.

Figure 1(d) shows our model's ring-and-hub structure. Although only mimicking the network geometry of naturally occurring fungi (Figs. 1(a) and (b)), it is actually a very realistic model for current experiments in both fungal and slime-mold systems. In particular, experiments have already been carried out with food-sources placed at peripheral nodes for fungi (Fig. 1(e)) and slime-mold \cite{slime} with the resulting network structures showing a wide range of complex morphologies \cite{slime}. We use the term  `hub' very generally, to describe a central portion of the network where many paths may pass but
where significant transport delays might arise. Such delays represent an effective cost for passing through the hub. In practice, this delay may be due to (i) direct congestion at some central junction, or (ii) transport through some central portion which itself has a network structure (e.g. the inner ring of the Houston road network  in Fig. 1(f)). We return to this point later on.

%%%%%%%%%%%%%%%%%%%%%%%%%% FIGURE 1 %%%%%%%%%%%%%%%%%%%
\begin{figure}[ht]
\includegraphics[width=.50\textwidth]{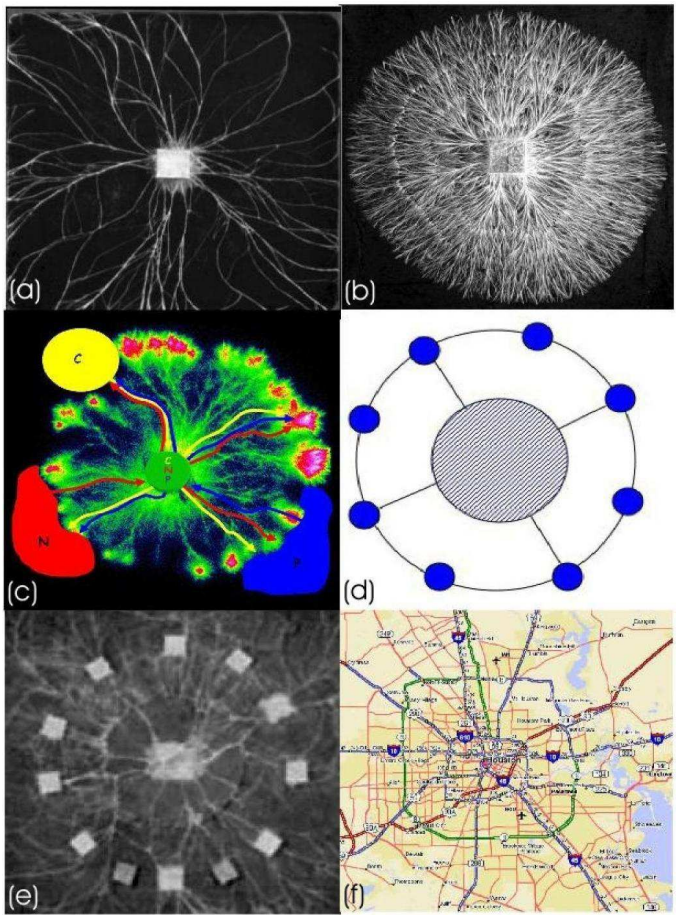}
\caption{
(a) Typical network for the fungus {\em Resinicium bicolour}.     
(b) Typical network for the fungus {\em Hypholoma fasciculare}. This network has a far denser set of  connections than (a), yet both are able to thrive in the same environmental conditions. (c) Schematic representation of the nutrient flows through a mycelial fungus. The food resources (carbon (C), nitrogen (N), phosphorous (P)) need to be transported efficiently across the entire structure.
(d) Our analytically-solvable model showing radial connections from peripheral nodes to an effective hub.
(e) Mycelial fungus {\em Phanerochaete velutina} after 98 days growing from a central hub-like resource. From day 48, the system is 
    supplied with pairs of fresh 4 cm$^3$ wood resources at 10 day intervals. The 
    resultant network has both radial and circumferential connections, as in our model (Fig. 1(d)).
(f) The man-made road network in Houston, showing a complicated inner `hub' which contains an
    embedded inner ring.}
\label{fig:figure1}
\end{figure}
%%%%%%%%%%%%%%%%%%%%%%%%%% FIGURE 1 %%%%%%%%%%%%%%%%%%%

Building on earlier work \cite{prl,dm}, we can calculate analytically the distribution of path lengths in a ring-and-hub structure in which $n$ peripheral nodes (e.g. food sources) are connected to the central hub portion with a probability $p$. The mean number of radial connections (i.e. anastomoses in the case of fungi) is equal to $n p$. Assuming that all the connections are continually transporting objects (e.g. nutrients) at the same speed, any time delay for passing through the center can be represented equivalently as an additional path-length. This additional path-length is given by $c$ which can be taken to be a general function of the system parameters. It can then be shown that the probability $P(\ell)$ that the shortest path between a pair of nodes on the perimeter is $\ell$, is given (when averaged over all pairs of nodes) by:
{\small \begin{displaymath}
P(\ell) = \left\{ \begin{array}{ll}
 \frac{1}{n-1} & \ \ \textrm{$\forall$ $\ell \leq c$}\\
 \frac{1}{n-1}\bigl[1+(\ell -c-1)p+ \\
\ \ \ \ \ \  (n-1-\ell )(\ell -c)p^2 \bigr](1-p)^{\ell-c-1} & 
\ \ \textrm{$\forall$  $\ell > c$\ .} \end{array} \right.
\end{displaymath}}
This expression is valid for $c<n$, since for $c\geq n$ the shortest path is trivially around the ring itself.
The average shortest path is now given by $\bar{\ell} = \sum_{\ell=1}^{n-1}\ell P(\ell)$. Performing the relevant summations yields:
{
\small
  \begin{eqnarray}\label{c-m-l}
\bar{\ell }(p,n,c)=\frac{(1-p)^{n-c}\bigl(3+(n-2-c)p\bigr)}{p^2(n-1)}
\ \ \ \ \ \ \ \ \ \ \ \ \ \ \ \ \ \ \ \ \ \ \ \nonumber\\
+\frac{p\bigl(2-2c+2n-(c-1)(c-n)p\bigr)-3}{p^2(n-1)}+\frac{c(c-1)}{2(n-1)}\ \ \ \ \ \ 
  \end{eqnarray}
}
which shows explicitly the parameter dependences of $\bar{\ell}$. The non-trivial algebraic structure of this expression shows that the congestion cost $c$ cannot be treated in a simple perturbation scheme -- in particular, the term $(1-p)^{n-c}$ represents an infinite summation of such perturbative terms.

We now consider the functional form of $\bar{\ell}$ for different cost
functions $c$. In Ref. \cite{prl} we examined the trivial cases of fixed, linear and quadratic costs. Here we show that for more general non-linear cost-functions, a novel and highly non-trivial phase-transition can arise.
First we consider the case of a general cubic cost function:
\begin{equation}
  c(\rho) = A\rho^3 + B\rho^2 + C\rho + D,
\end{equation}
where $\rho$ is the scaled probability, $\rho = pn$ and $A, B, C,
D\in\mathbf{R}$. In order to demonstrate the richness of the phase transition and yet still keep a physically reasonable model, we choose
the minimum in this cubic function to be a stationary point. Hence
the cost function remains a monotonically increasing function, but features a regime of intermediate connectivities over which congestion costs remain essentially flat (like the `fixed charge' for London's congestion zone).  Since we are primarily concerned with an
optimization problem, we can set the constant $D=0$. Hence
\begin{equation}\label{cubic-cost}
  c(\rho) = A\rho^3 - 3Ar\rho^2 + 3Ar^2\rho,
\end{equation}
where $r=\frac{-B}{3A}$ is the location of the stationary point.  Substituting into Eq. (\ref{c-m-l}) yields the
shortest path distribution for this particular cost-function in
terms of the parameters $A$, $r$, $p$ and $n$.  The result is too cumbersome to give explicitly -- however we emphasize that it is straightforward to obtain, it is exact, and it allows various limiting cases to be analyzed analytically.

%%%%%%%%%%%%%%%%%%%%%%%%%% FIGURE 2 %%%%%%%%%%%%%%%%%%%
\begin{figure}[ht]
\includegraphics[width=.48\textwidth]{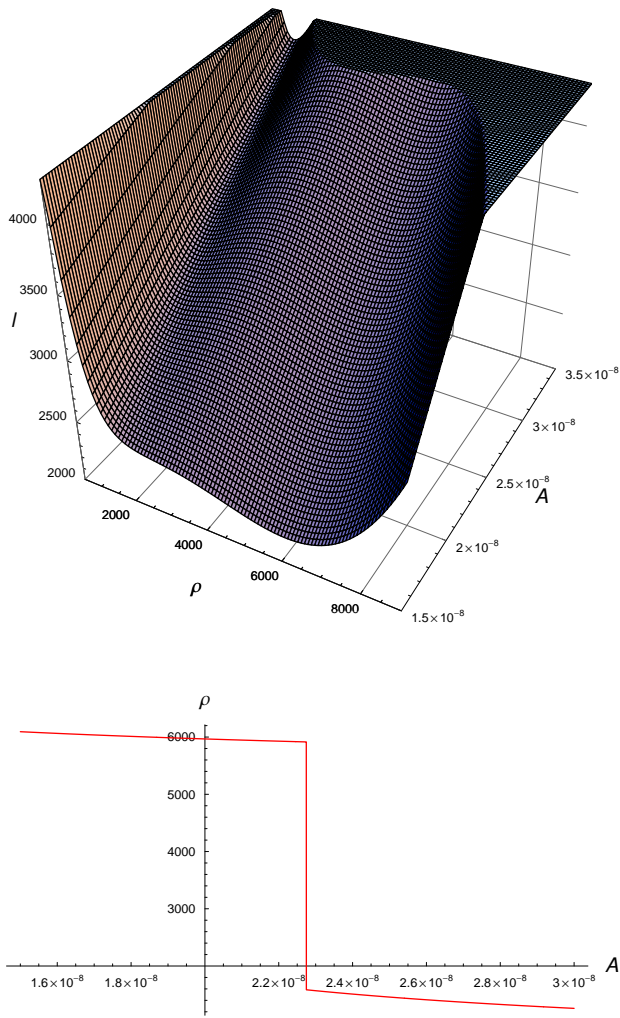}
\caption{Top: Landscape of the average shortest path length
  $\bar{\ell}$ (vertical axis) as a function of the cubic cost-function parameter $A$ and the
    average number of connections to the central hub $\rho$.  Bottom: The value of
    $\rho$ corresponding to a global minima in $\bar{\ell}$, as a function of the
    cubic cost-parameter $A$.}
\label{fig:figure2}
\end{figure}
%%%%%%%%%%%%%%%%%%%%%%%%%% FIGURE 2  %%%%%%%%%%%%%%%%%%%

Figure 2 (top) shows the value of the average shortest path $\bar{\ell}$ for varying values of $\rho$ and $A$.  
As can be seen, the optimal network structure (i.e. the network whose connectivity $\rho$ is such that $\bar{\ell}$ is a global minimum) changes abruptly from a high connectivity structure to a low connectivity one, as the cost-function parameter $A$ increases.  Figure 2 (bottom) shows that this transition resembles a first-order phase transition. At the transition point $A=A_{\rm crit}$, both the high and low connectivity structures are optimal. Hence there are two structurally inequivalent networks having identical (and optimal) functional properties. As we move below or above the transition point (i.e. $A<A_{\rm crit}$ or $A>A_{\rm crit}$ respectively) the high or low connectivity structure becomes  increasingly superior. 

%%%%%%%%%%%%%%%%%%%%%%%%%% FIGURE 3 %%%%%%%%%%%%%%%%%%%
\begin{figure}[ht]
\includegraphics[width=.48\textwidth]{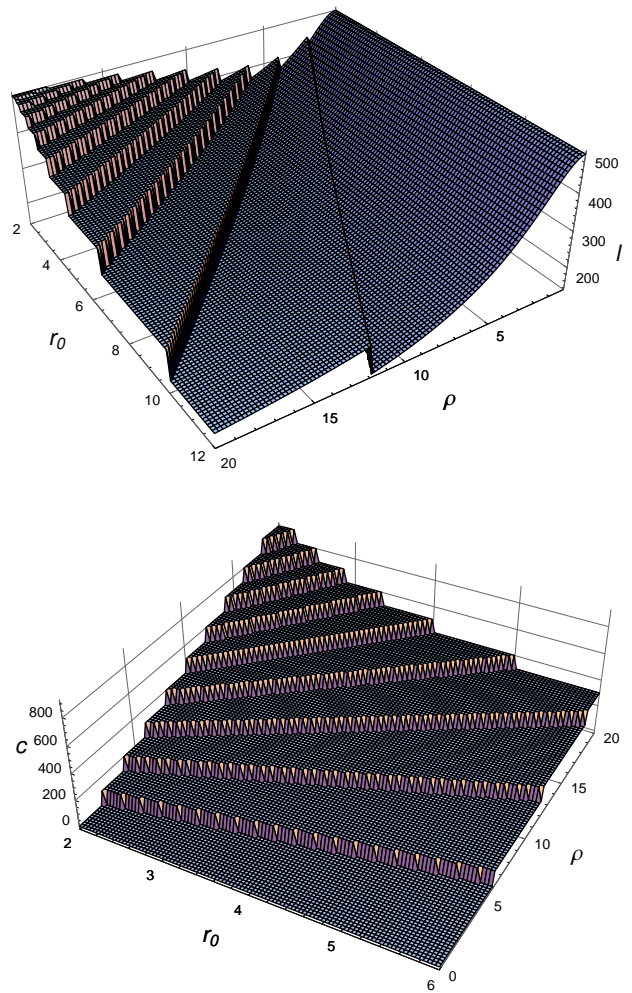}
\caption{ Top: Landscape of the average shortest path length
  $\bar{\ell}$ (vertical axis) as a function of the
`step' cost-function parameter $r_0$ and the
    connectivity $\rho$. Bottom:
The `step' cost-function as a function of the step-frequency parameter $r_0$ and $\rho$.  As $r_0$ decreases, the cost-function becomes increasingly 
linear. } \label{fig:figure3}
\end{figure}
%%%%%%%%%%%%%%%%%%%%%%%%%% FIGURE 3 %%%%%%%%%%%%%%%%%%%

%%%%%%%%%%%%%%%%%%%%%%%%%% FIGURE 4 %%%%%%%%%%%%%%%%%%%
\begin{figure}[ht]
\includegraphics[width=.5\textwidth]{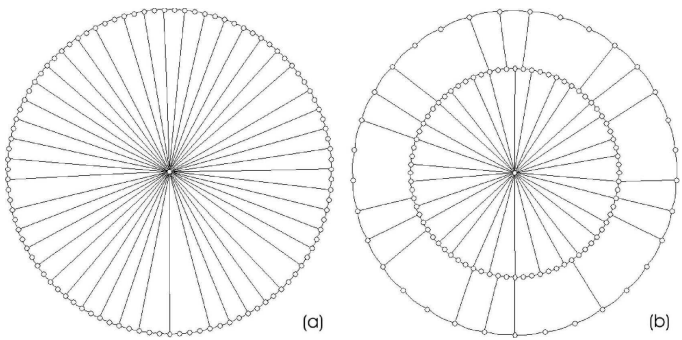}
\caption{Two structurally inequivalent but functionally equivalent networks. The average shortest path across the structure is
    the {\em same} for both networks, and both networks are themselves optimal (i.e. minimum $\bar{\ell}$). (a) A single-ring-and-hub network with a linear cost function.   (b) A two-ring-and-hub configuration. The inner
    ring-hub structure has the same cost-function as in (a). The similarity to the real fungi in Figs. 1(a) and (b) is striking.} \label{fig:figure4}
\end{figure}
%%%%%%%%%%%%%%%%%%%%%%%%%% FIGURE 4 %%%%%%%%%%%%%%%%%%%

We have checked that similar structural transitions can arise for higher-order nonlinear cost functions. In particular we  demonstrate here the extreme case of a `step' function, where the cost is fixed until the connectivity to the central hub portion reaches a particular threshold value. As an illustration, we consider the particular case: 
\begin{equation}\label{step-cost}
c(\rho,r_0) = 50 \bigg(\sum_{i=1}^{50} {\rm Sgn} (\rho - i r_0) + 50\bigg),
\end{equation}
where ${\rm Sgn}(x)=-1,0,1$ depending on whether $x$ is negative,
zero, or positive respectively, and $r_0$ determines the frequency of
the jump in the cost.  Figure 3 (top) shows the average shortest path $\bar{\ell}$
for this step cost-function (Fig. 3 (bottom)) as $\rho$ and $r_0$ are varied. 
A multitude of structurally-distinct yet optimal network configurations emerge.  
As $r_0$ decreases, the step-size in the cost function decreases and the cost-function itself begins to take on a  linear form -- accordingly, the behavior of $\bar{\ell}$ tends towards that of a linear cost model
with a single identifiable minimum \cite{prl}.
Most importantly, we can see that once again a
gradual change in the cost parameter leads to an abrupt change in the
structure of the optimal (i.e. minimum $\bar{\ell}$) network.

We have allowed our ring-and-hub networks to seek optimality by modifying their radial connectivity while maintaining a single ring. Relaxing this constraint to allow for transitions to multiple-ring structures yields similar findings.  In particular, allowing both the radial connectivity and the number of rings to change yields abrupt transitions between optimal networks with different radial connectivities {\em and} different numbers of rings. To analyze analytically such multiple-ring structures we introduce the following renormalization scheme. Consider the two-ring-and-hub network in Fig. 1(f). For paths which pass near the center, there is a contribution to the path-length resulting from the fact that the inner ring has a network structure which needs to be traversed. Hence the inner-ring-plus-hub portion acts as a {\em renormalized hub} for the outer ring. In short, the ring-plus-hub of Eq. (1) can be treated as a `cost motif' for solving multiple-ring-and-hub problems, by allowing us to write a recurrence relation which relates the average shortest path in a network with $i+1$ rings to that for $i$ rings:
{
\small
  \begin{eqnarray}\label{c-m-l-rings}
\bar{\ell }_{i+1}(p_{i+1},n_{i+1},c)= 
\ \ \ \ \ \ \ \ \ \ \ \ \ \ \ \ \ \ \ \ \ \ \ \ \ \ \ \ \ \ \ \ \ \ \ \ \ \ \ \
\ \ \ \ \ \ \ \ \ \ \ \ \ \ \ \ \ \ \ \ \ \ \ \ \ \ \ \ \ \ \ \ \ \ \ \ \ \ \ \ 
\ \ \ \ \ \ \ \ \ \ \ \ \ \ \ \ \ \ \ \ \ \ \ \ \ \ \ \ \ \ \ \ \ \ \ \ \ \ \nonumber\\
\frac{(1-p_{i+1})^{n_{i+1}-\bar{\ell }_{i}(p_{i},n_{i},c)}\bigl(3+(n_{i+1}-2-\bar{\ell }_{i}(p_{i},n_{i},c))p_{i+1}\bigr)}{p_{i+1}^2(n_{i+1}-1)}
\ \ \ \ \ \ \ \ \ \ \ \ \ \ \ \ \ \ \ \ \ \ \ \ \ \ \ \ \ \ \ \ \ \ \ \ \ \ \ \
\ \ \ \ \ \ \ \ \ \ \ \ \ \ \ \ \ \ \ \ \ \ \ \ \ \ \ \ \ \ \ \ \ \ \ \ \ \ \ \
\ \ \nonumber\\
+\frac{2p_{i+1}\bigl(1-\bar{\ell }_{i}(p_{i},n_{i},c)+n_{i+1})}{p_{i+1}^2(n_{i+1}-1)}
\ \ \ \ \ \ \ \ \ \ \ \ \ \ \ \ \ \ \ \ \ \ \ \ \ \ \ \ \ \ \ \ \ \ \ \ \ \ \ \
\ \ \ \ \ \ \ \ \ \ \ \ \ \ \ \ \ \ \ \ \ \ \ \ \ \ \ \ \ \ \ \ \ \ \ \ \ \ \ \
\ \ \ \ \ \ \ \ \ \ \ \ \ \ \ \nonumber\\
-\frac{p_{i+1}\bigl((\bar{\ell }_{i}(p_{i},n_{i},c)-1)(\bar{\ell }_{i}(p_{i},n_{i},c)-n_{i+1})p_{i+1}\bigr)-3}{p_{i+1}^2(n_{i+1}-1)}
\ \ \ \ \ \ \ \ \ \ \ \ \ \ \ \ \ \ \ \ \ \ \ \ \ \ \ \ \ \ \ \ \ \ \ \ \ \ \ \
\ \ \ \ \ \ \ \ \ \ \ \ \ \ \ \ \ \ \ \ \ \ \ \ \ \ \ \ \ \ \ \ \ \ \ \ \ \ \ \
\ \ \ \ \ \ \ \ \ \ \ \ \ \ \ \nonumber\\
+\frac{\bar{\ell }_{i}(p_{i},n_{i},c)(\bar{\ell }_{i}(p_{i},n_{i},c)-1)}{2(n_{i+1}-1)}\, \
\ \ \ \ \ \ \ \ \ \ \ \ \ \ \ \ \ \ \ \ \ \ \ \ \ \ \ \ \ \ \ \ \ \ \ \ \ \ \ \
\ \ \ \ \ \ \ \ \ \ \ \ \ \ \ \ \ \ \ \ \ \ \ \ \ \ \ \ \ \ \ \ \ \ \ \ \ \ \ \
\ \ \ \ \ \ \ \ \ \ \ \ \ \ \
\end{eqnarray}
}
 where $i \geq 0$ and $\bar{\ell}_{0} = c$ with $c$ being a general cost for the inner-most hub. The case $i=0$ is identical to Eq. (1). As before, $p_{i+1}$
represents the probability of a link between rings $i+1$ and $i$ and $n_{i+1}$ is the
number of nodes in ring $i+1$.  We have checked that this recurrence relation is accurate, using full-scale numerical calculations of $\bar{\ell}$. Solving this recurrence relation shows that there
are optimal network structures with different numbers of rings and radial connectivities, yet which have the {\em same} average shortest path length across them \cite{note}. Hence, as before, optimal network structures exist which are structurally very different, yet functionally equivalent. Figure 4 shows an explicit example of two such functionally equivalent, optimal networks. It is remarkable that these images are so similar to the real fungi shown in Figs. 1(a) and (b).

In summary, we have uncovered a novel structural phase transition within a class
of biologically-motivated networks. Depending on the system of interest (e.g.
    fungus, or road networks) these transitions between inequivalent structures
might be explored in real-time by adaptive re-wiring, or over successive
generations through evolutionary forces or `experience'.  An important further
implication of this work is that in addition to searching for a universality in
terms of network structure, one might fruitfully consider seeking universality in
terms of network {\em function}.

We kindly acknowledge L. Boddy, J. Wells, M. Harris and G. Tordoff (University of Cardiff) for the fungal images in Figs. 1(a), (b) and (e). N.J. and M.F. acknowledge the support of EU through the  MMCOMNET project.


\begin{thebibliography}{99}

\bibitem{newman} M.E.J. Newman, SIAM Review {\bf 45}, 167 (2003).

\bibitem{newman2} M.T. Gastner, M.E.J. Newman, cond-mat/0409702. This paper looks at shape and efficiency from a different viewpoint -- effects of transport congestion are not included.

\bibitem{watts98} D.J. Watts and S.H. Strogatz, Nature {\bf 393}, 440 (1998).

\bibitem{nets} D. S. Callaway, M. E. J.  Newman, S. H.  Strogatz, and D. J.
Watts, Phys. Rev. Lett. {\bf 85}, 5468 (2000)

\bibitem{nets2} R. Albert and A.L.  Barabasi, Phys. Rev. Lett. {\bf 85}, 5234
(2000).

\bibitem{charges} L.A. Brunstein, S.V. Buldyrev, R. Cohen, S. Havlin and H.E.
Stanley, Phys. Rev. Lett. {\bf 91}, 168701 (2003).

\bibitem{central} 
R. Guimera, A. Diaz-Guilera et al., Phys. Rev. Lett. {\bf 89}, 248701 (2002).

\bibitem{search} V. Colizza, J. R. Banavar et al., Phys. Rev. Lett. {\bf 92}, 
198701 (2004).

\bibitem{gradient} Z. Toroczkai, K. E. Bassler, Nature {\bf 428}, 716 (2004).

\bibitem{prl} D.J. Ashton, T.C. Jarrett, and N.F. Johnson, Phys. Rev. Lett. {\bf 94}, 058701 (2005).

\bibitem{dm} S.N. Dorogovtsev and J.F.F. Mendes, Europhys. Lett. {\bf 50}, 1
(2000).

\bibitem{mark} M. Tlalka, D. Hensman, P.R. Darrah, S.C. Watkinson and M.D.
Fricker, New Phytologist {\bf 158}, 325 (2003).

\bibitem{slime} T. Nakagaki, R. Kobayashi, Y. Nishiura, and T. Ueda, Proc. R. Soc. Lond. B {\bf 271}, 2305 (2004).

\bibitem{note} The networks need to be optimal in order that this question of equivalence be meaningful and non-trivial. By contrast, it is a fairly trivial exercise to find {\em non-optimal} structures which are functionally equivalent. 


\end{thebibliography}
\end{document}